\documentclass[aps,pra,twocolumn,reprint,groupedaddress,noshowpacs,floatfix]{revtex4-1}

\usepackage{bm} 
\usepackage{color}
\usepackage{amsmath}
\usepackage{amssymb}
\usepackage{mathrsfs}
\usepackage{graphicx} 
\usepackage{lastpage}
\usepackage{epstopdf}
\usepackage[titletoc,title]{appendix}
\usepackage[normalem]{ulem} 

\newcommand{\be}{\begin{equation}}
\newcommand{\ee}{\end{equation}}
\newcommand{\ba}{\begin{eqnarray}}
\newcommand{\ea}{\end{eqnarray}}
\newcommand{\re}{\mathrm{e}}            

\newcommand{\bcalF}{\mbox{\boldmath$\cal F$}}
\newcommand{\bcalG}{\mbox{\boldmath$\cal G$}}

\newcommand{\bCh}{\mathrm {\bf Ch}}

\begin{document}

\title{Unidirectional propulsion of planar magnetic nanomachines}

\author{Kevin-Joshua Cohen$^1$}
\author{Boris Y. Rubinstein$^2$}
\author{Oded Kenneth$^3$}
\author{Alexander M. Leshansky$^4$}
\email{lisha@technion.ac.il}\affiliation{$^1$Department of Mathematics, Technion -- Israel Institute of Technology, Haifa 32000, Israel\\
$^2$Stowers Institute for Medical Research, Kansas City, MO 64110, USA\\
$^3$Department of Physics, Technion -- Israel Institute of Technology, Haifa, 32000, Israel\\
$^4$Department of Chemical Engineering, Technion -- Israel Institute of Technology, Haifa 32000, Israel}

\date{\today}

\begin{abstract}
Steering of magnetic nano-/microhelices by a rotating magnetic field is considered as a promising technique for controlled navigation of tiny objects through viscous fluidic environments. It has been recently demonstrated that simple geometrically achiral planar structures can also be steered efficiently. Such planar propellers are interesting for practical reasons, as  they can be mass-fabricated using standard micro/nanolithography techniques. While planar magnetic structures are prone to in-plane magnetization,
under the effect of an in-plane rotating magnetic field, they exhibit, at most, propulsion due to spontaneous symmetry breaking, i.e., they can propel either parallel or anti-parallel to the rotation axis of the field depending on their initial orientation. Here we demonstrate that actuation by a \emph{conically} rotating magnetic field
(i.e., superposition of in-plane rotating field and constant field orthogonal to it) can yield efficient \emph{unidirectional} propulsion of planar and magnetized in-plane structures. In particular, we found that a highly symmetrical V-shape magnetized along its symmetry axis which exhibits no net propulsion in in-plane rotating field, exhibits unidirectional in-sync propulsion with a \emph{constant} (frequency-independent) velocity when actuated by the conically rotating field.

\end{abstract}

\maketitle

\section{Introduction}
Controlled propulsion of artificial micro- and nano-structures that can be actuated and precisely navigated through a fluidic environment has recently attracted considerable attention. While many different approaches ranging from catalytic nanowires to thermally, light- and acoustically-driven nanomachines are being explored, driven propulsion powered by an external \emph{rotating} magnetic field offering remote, engine-less and fuel-free steering of micro-/nanostructures, is particularly appealing for prospective biomedical applications (see \cite{wang,afm18} for review).

Originally bio-inspired helical micro/nanopropellers were demonstrated \cite{GF,Nelson09} and extensively studied, e.g., \cite{Ghosh13,ML14a,ML14b,Walker}. Such helical `swimmers' are actuated by the weak (few milli Tesla) uniform in-plane rotating magnetic field and propel unidirectionally along the field rotation axis similar to a twirling bacterial flagella. However, fabrication of three-dimensional (3D) helical micro- and nanoscale structures typically requires complicated fabrication techniques, e.g, ``top-down" approach \cite{Nelson09}, glancing angle deposition \cite{GF,Walker}, direct laser writing \cite{Tot}, biotemplated synthesis using biological spiral structures \cite{xylem,spirulina}, two-photon polymerization of a curable superparamagnetic polymer composite \cite{PetersIEEE13,Peters15}, spiraling microfluidic flow lithography \cite{flowlit18}, etc. One interesting proposal to circumvent complicated microfabrication is to use one-dimensional \emph{soft} magnetic nanowires \cite{Gao,Pak} that supposedly acquire helicity when actuated by rotating field due to an interplay of viscous and elastic forces. An alternative method that does not require sophisticated microfabrication involves spontaneous aggregation of magnetic nanoparticles into random-shaped 3D clusters \cite{Vach1,Vach2}. However, on average such random-shaped clusters appear to be significantly less efficient propellers in comparison to the structures with preprogrammed geometry and magnetization \cite{scirob}.

Another interesting option relies on the fact that geometric chirality is not required for driven propulsion based on rotation-translation coupling. It was recently demonstrated that geometrically \emph{achiral} planar objects made of three interconnected magnetized microbeads can be steered quite efficiently by an in-plane rotating magnetic field \cite{Cheang,ASME14}. Such two-dimensional (2D) ferromagnetic propellers are of practical interest, as they can be mass-fabricated via standard photolithography methods \cite{tottori18}. Recently developed microfluidic stop-flow lithography can also be used for high-throughput fabrication of superparamagnetic 2D microstructures with high saturation magnetization \cite{stopflow}.

\begin{figure}
\centering
\includegraphics[scale=1.1]{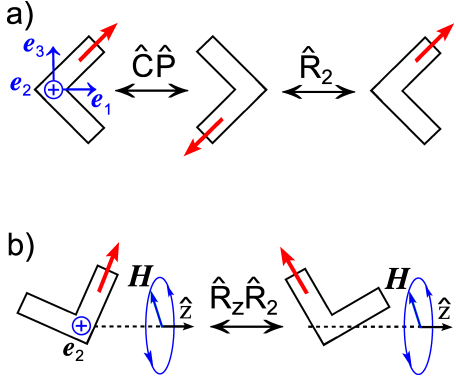}
\caption{Schematic diagrams illustrating (a) $\widehat{C}\widehat{P}$-symmetry of a slim planar V-shape propeller magnetized in its plane, (b) A pair of dual solutions related by symmetry and having opposite propulsion velocities along $+z$ (left) and $-z$ (right) direction, respectively, when actuated by an in-plane rotating magnetic field, ${\bm H}$. Principal axes of rotation
$\{\bm{e}_1, \bm{e}_2, \bm{e}_3\}$ are shown in (a); red arrow stands for the magnetic dipole moment $\bm m$; $\widehat{R}_2$ stands for  rotation by $\pi$ around the body frame principal axis ${\bm e}_2$.
\label{fig:symmetry} }
\end{figure}

The theory of magnetically driven propulsion of an \emph{arbitrary} shaped object was developed in \cite{prf17}, suggesting that the notion of \emph{chirality} should account not just for the object's geometry, but also for orientation of the magnetic dipolar moment affixed to it. In particular it was predicted that specific magnetization of the geometrically achiral planar object can actually render it chiral resulting in unidirectional propulsion similar to helices. A combined theoretical and experimental study of planar V-shaped structures actuated by an in-plane rotating magnetic (or electric) field was recently conducted in Ref.~\cite{pre18}. The correspondence (depending on orientation of the dipolar moment) between different propulsive solutions was established based on symmetry arguments involving parity, $\widehat{P}$, and charge conjugation, $\widehat{C}$, as the in-plane rotating magnetic field is invariant under $\widehat{P}$ and $\widehat{C}\widehat{R}_z$. Here $\widehat{R}_z$ stands for rotation by $\pi$ around
the field rotation $z$-axis. In general, there could be two stable rotational solutions resulting in different propulsion velocities. In particular, it was found that highly symmetrical achiral ($\widehat{P}$-even) V-shaped objects (e.g., with magnetization along $\bm e_3$ or $\bm e_2$, see Fig.~\ref{fig:symmetry}a) exhibit no net propulsion at all. Furthermore, it was demonstrated that a V-shaped object, magnetized along any \emph{principal axis of rotation} will exhibit no net in-sync propulsion regardless of its symmetry \cite{pre18}. Individual magnetized in-plane (as in Fig.~\ref{fig:symmetry}a) $\widehat{C}\widehat{P}$-even objects can efficiently propel due to \emph{spontaneous} symmetry breaking. However, since $\widehat{C}\widehat{P}$-symmetry inverts linear velocities, the dual rotational solutions yield propulsion with equal, but opposite velocities (see illustration in Fig.~\ref{fig:symmetry}b). It was also confirmed experimentally that in agreement with the prediction in \cite{prf17} the off-plane magnetized V-shaped object with intrinsically broken ($\widehat{C}\widehat{P}$-and $\widehat{P}$-) symmetry, i.e., chiral as well as $\widehat{C}\widehat{P}$-chiral, can propel unidirectionally.

Since planar micro/nano-structures are prone to \emph{in-plane} magnetization while uniform off-plane magnetization of multiple samples is not an easy task, the interesting question is whether planar nanopropellers can be steered in a controllable fashion? As it will be shown below, controlled propulsion can be achieved using a \emph{conically} rotating field, i.e., by adding an extra constant magnetic field along the field rotation $z$-axis breaking the $\widehat{C}\widehat{R}_z$-symmetry. The role of the constant field is to orient the magnetic moment along the $z$-axis and this results in selection of one of the dual solutions in Fig.~\ref{fig:symmetry}b (on the left) over the other.
We also shall demonstrate that a highly symmetrical planar V-object magnetized along its symmetry axis ($\bm e_1$-axis in Fig.~\ref{fig:symmetry}a) which shows no net propulsion in an in-plane rotating field (see \cite{pre18}), can be steered \emph{unidirectionally} by the conically rotating field similar to a magnetic helix.

\section{Problem formulation \label{sec:general}}
We assume \emph{conical} rotating magnetic field $\bm {H}$
\be
\bm {H}=H(\hat{\bm x}\cos{\omega t}+ \hat{\bm y}\sin{\omega t} +\hat{\bm z} \delta)\,, \label{eq:field}
\ee
where $H$ and $\omega$ are, respectively, the amplitude and angular frequency of the \emph{rotating} field and $H_z=\delta H$ is the value of the constant magnetic field along the field rotation $z$-axis, such that $\tan^{-1}{(1/\delta)}$ is the cone angle.

We further assume that the motion of the magnetized object is force--free and driven solely by the magnetic torque $\bm{L}={\bm m}\times\bm{H}$, where $\bm m$ is the magnetic moment affixed to the object. In the zero-Reynolds-number (Stokes) approximation, the condition of the balance of forces and torques
acting on the particle reads
\be
{\bm U}={\bcalG}\cdot \bm{L}\,, \qquad
\bm{\mathit{\Omega}}={\bcalF}\cdot \bm{L}\,. \label{eq:U}
\ee
Here $\bm{U}$ and $\bm{\mathit{\Omega}}$ are the translational and angular velocities of body,
${\bcalG}$ and ${\bcalF}$ are the coupling and rotation viscous mobility tensors, respectively.
The triad of unit eigenvectors,  $\left\{{\bm e}_1, {\bm e}_2, {\bm e}_3\right\}$ of ${\bcalF}$
makes up the body-frame \emph{principal rotation axes}. We fix their order such that the corresponding eigenvalues satisfy
${\mathcal F}_{1}\le{\mathcal F}_{2}\le{\mathcal F}_{3}$. The lab-frame unit vectors $\{\hat{\bm x},\hat{\bm y},\hat{\bm z}\}$ are related to the body-frame axes $\{\bm e_1, \bm e_2, \bm e_3\}$ by a rotation matrix ${\bm R}$ parameterized by, e.g., the three Euler angles $\varphi$, $\theta$ and $\psi$ (standard ``3-1-3" parametrization) describing the instantaneous orientation of the object in the lab frame,
\ba
{\bm R} &=&
\begin{pmatrix}
c_{\psi} & s_{\psi} &  0 \\
-s_{\varphi} & c_{\psi} &  0 \\
0 & 0 &  1
\end{pmatrix}
\cdot
\begin{pmatrix}
1 & 0 &  0 \\
0 & c_{\theta} &  s_{\theta} \\
0 & -s_{\theta} &  c_{\theta}
\end{pmatrix}
\cdot
\begin{pmatrix}
c_{\varphi} & s_{\varphi} &  0 \\
-s_{\varphi} & c_{\varphi} &  0 \\
0 & 0 &  1
\end{pmatrix}= \nonumber \\
&& \begin{pmatrix}
c_{\varphi}c_{\psi}-s_{\varphi}s_{\psi}c_{\theta} & s_{\varphi}c_{\psi}+c_{\varphi}s_{\psi}c_{\theta} &  s_{\psi}s_{\theta} \\
-c_{\varphi}s_{\psi}-s_{\varphi}c_{\psi}c_{\theta} & -s_{\varphi}s_{\psi}+c_{\varphi}c_{\psi}c_{\theta} &  c_{\psi}s_{\theta} \\
s_{\varphi}s_{\theta} & -c_{\varphi}s_{\theta} &  c_{\theta}
\end{pmatrix}\,,\label{eq:R}
\ea
where we use the compact notation: $s_{\theta}\equiv\sin{\theta}$, $c_{\psi}\equiv\cos{\psi}$, etc. For an arbitrary vector $\bm A$ we have
$\bm A^b={\bm R}\cdot {\bm A}^l$ (or $\bm A^l={\bm R}^T\cdot {\bm A}^b$), where superscripts ``$b$" and ``$l$" stand for the body- and lab-frame of reference respectively.

The permanent magnetic moment in the body-frame axes is given by
\be
{\bm m}=m\,(n_1 \bm e_1 + n_2 \bm e_2 + n_3 \bm e_3)\,, \label{eq:moment}
\ee
where $n_i$ are projections of the unit vector ${\bm n}={\bm m}/m=s_\Phi c_\alpha \bm e_1 + s_\Phi s_\alpha \bm e_2 + c_\Phi \bm e_3$ expressed via the spherical polar, $\Phi$, and azimuthal, $\alpha$,  angles, respectively.\\

\noindent\emph{Angular velocity}. It is most convenient to express the problem of driven rotation (the second equation in (\ref{eq:U})) in the body-frame where ${\bcalF}=\mathrm{diag}(\mathcal F_1, \mathcal F_2, \mathcal F_3)$ is fixed and the components of the angular velocity $\bm{\mathit{\Omega}}$ are expressed through the Euler angles via the relations \cite{prf17}:
\be
{\mathit{\Omega}}_1^b=\dot{\varphi}s_{\theta}s_{\psi}+\dot{\theta}c_{\psi}, \;{\mathit{\Omega}}_2^b=\dot{\varphi}s_{\theta}c_{\psi}-\dot{\theta}s_{\psi}, \;{\mathit{\Omega}}_3^b=\dot{\varphi}c_{\theta}+\dot{\psi}, \label{eq:Omega}
\ee
where the dot stands for the time derivative. Expressing the magnetic field (\ref{eq:field}) in the body-frame components, $\bm H^b=\bm R \cdot \bm H^l$,
the equation $\bm{\mathit{\Omega}}={\bcalF} \cdot ({\bm m}\times {\bm H})$ after some algebra reduces to
\ba
&&\textstyle\frac{1+\varepsilon}{\omega_0}(\dot{\varphi}s_{\theta}s_{\psi}+\dot{\theta}c_{\psi})  = n_2  \zeta +
n_3 [c_{\widehat{\varphi}}s_{\psi}+ \chi c_{\psi}],  \label{eq:1} \\
&&\textstyle\frac{1-\varepsilon}{\omega_0} (\dot{\varphi}s_{\theta}c_{\psi}-\dot{\theta}s_{\psi}) = -n_1 \zeta +
n_3 [c_{\widehat{\varphi}}c_{\psi}- \chi s_{\psi}],   \label{eq:2} \\
&&\textstyle\frac{1}{p\omega_0} (\dot{\varphi}c_{\theta}+\dot{\psi})=-n_{\perp}[c_{\widehat{\varphi}} s_{\psi+\alpha} + \chi c_{\psi+\alpha}],  \label{eq:3}
\ea
where we denote $\zeta=s_{\theta}s_{\widehat{\varphi}} + \delta c_\theta$ and $\chi=s_{\widehat{\varphi}}c_{\theta}-\delta s_\theta$. Here ${\widehat{\varphi}}=\varphi - \omega t$, $n_{\perp}=\sin{\Phi}$,  $\omega_0=mH {\cal F}_{\perp}$ the characteristic angular frequency with ${\cal F}_{\perp}$ being the geometric mean minor mobilities, ${\cal F}_{\perp}^{-1} = ({\cal F}_1^{-1}+{\cal F}_2^{-1})/2$ and  $p={\cal F}_3/{\cal F}_{\perp} \ge 1$ and $\varepsilon=({\cal F}_2-{\cal F}_1)/({\cal F}_2+{\cal F}_1)\ge 0$ are, respectively, the \emph{longitudinal} and the \emph{transverse} rotational anisotropy parameters. For the in-plane rotating field ($\delta=0$) the Eqs.~(\ref{eq:1}-\ref{eq:3}) reduce to Eqs.(5-7) in \cite{prf17}.

We look for solutions which turn in-sync with the magnetic field, i.e., rotating about the $z$-axis with angular velocity
\be
\bm{\mathit{\Omega}}={\omega}\hat{z}=\omega(s_{\theta}s_{\psi}{\bm e}_1+ s_{\theta}c_{\psi}{\bm e}_2+ c_{\theta}{\bm e}_3)\,,\label{eq:Omega1}
\ee
From the comparison of (\ref{eq:Omega1}) and (\ref{eq:Omega}) it follows that the in-sync regime corresponds to constant values of the three Euler angles
$\psi$, $\theta$ and $\widehat{\varphi}=\varphi-\omega t$, so that Eqs.~\ref{eq:3} simplify to
\ba
&&(1+\varepsilon){\widetilde\omega} s_{\theta}s_{\psi}=n_2  \zeta +
n_3 [c_{\widehat{\varphi}}s_{\psi}+ \chi c_{\psi}],\;  \label{eq:1b} \\
&&(1-\varepsilon){\widetilde\omega} s_{\theta}c_{\psi}=-n_1 \zeta +
n_3 [c_{\widehat{\varphi}}c_{\psi}- \chi s_{\psi}] \; \label{eq:2b} \\
&&p^{-1}{\widetilde\omega} c_{\theta}=-n_{\perp}[c_{\widehat{\varphi}} s_{\psi+\alpha} + \chi c_{\psi+\alpha}] \;  \label{eq:3b}
\ea
where $\widetilde{\omega}=\omega/\omega_0$ is the dimensionless actuation frequency.

The Eqs.~(\ref{eq:1b})--(\ref{eq:3b}) shall be used to make analytical progress in some particular cases, e.g., magnetization along principal rotation axes, where one may expect net propulsion (in comparison with the case of $\delta=0$ where net propulsion is not possible \cite{pre18}). In general numerical solution of (\ref{eq:1b})--(\ref{eq:3b}) or time integration of (\ref{eq:1})--(\ref{eq:3}) for some initial values of the angles is required.\\

\noindent\emph{Linear velocity}. The linear velocity, $\bm U$ can be found in the same way as was done for in-plane rotating field \cite{prf17,scirob,pre18}, as the constant
component of the field $\delta H \hat{\bm z}$ affects propulsion only through dynamic orientation of the propeller. Expressing the magnetic torque, $\bm{L}$, from the second equation in (\ref{eq:U}) and substituting it into the the first equation in (\ref{eq:U}), the translational velocity can be readily found as ${\bm U}={\bcalG}\cdot{\bcalF}^{-1}\cdot \bm{\mathit{\Omega}}$. By symmetry the time-averaged linear velocity for in-sync actuation is along the $z$-axis. Taking a scalar product on both sides of this equation with $\bm{\mathit{\Omega}}=\omega \hat{\bm z}$ we readily obtain it in a compact covariant form as
\begin{equation}
\frac{U_z}{\omega\ell}=\widehat{\bm{\mathit{\Omega}}}\cdot\bCh
\cdot\widehat{\bm{\mathit{\Omega}}}\;,\label{eq:Uz}
\end{equation}
where $\bCh$ is a dimensionless chirality matrix given by the symmetric part of $\frac{1}{\ell}\, \bcalG \cdot \bcalF^{-1}$ with $\ell$
being the characteristic length and $\widehat{\bm{\mathit{\Omega}}}=\bm{\mathit{\Omega}}/\omega=\hat{\bm z}$ the normalized (unit) angular velocity.
It  is most convenient to write the RHS of (\ref{eq:Uz}) in the body frame whereas $\bCh$ is fixed and $\widehat{\bm{\mathit{\Omega}}}$ being expressed via the Euler angles as in (\ref{eq:Omega1}). Note that $\bCh$ (in contrast to $\bcalG$) is independent of the choice of coordinate origin. Under rotation of the coordinate frame it transforms as a (symmetric) pseudo-tensor.

\begin{figure} \centering
\includegraphics[scale=.2]{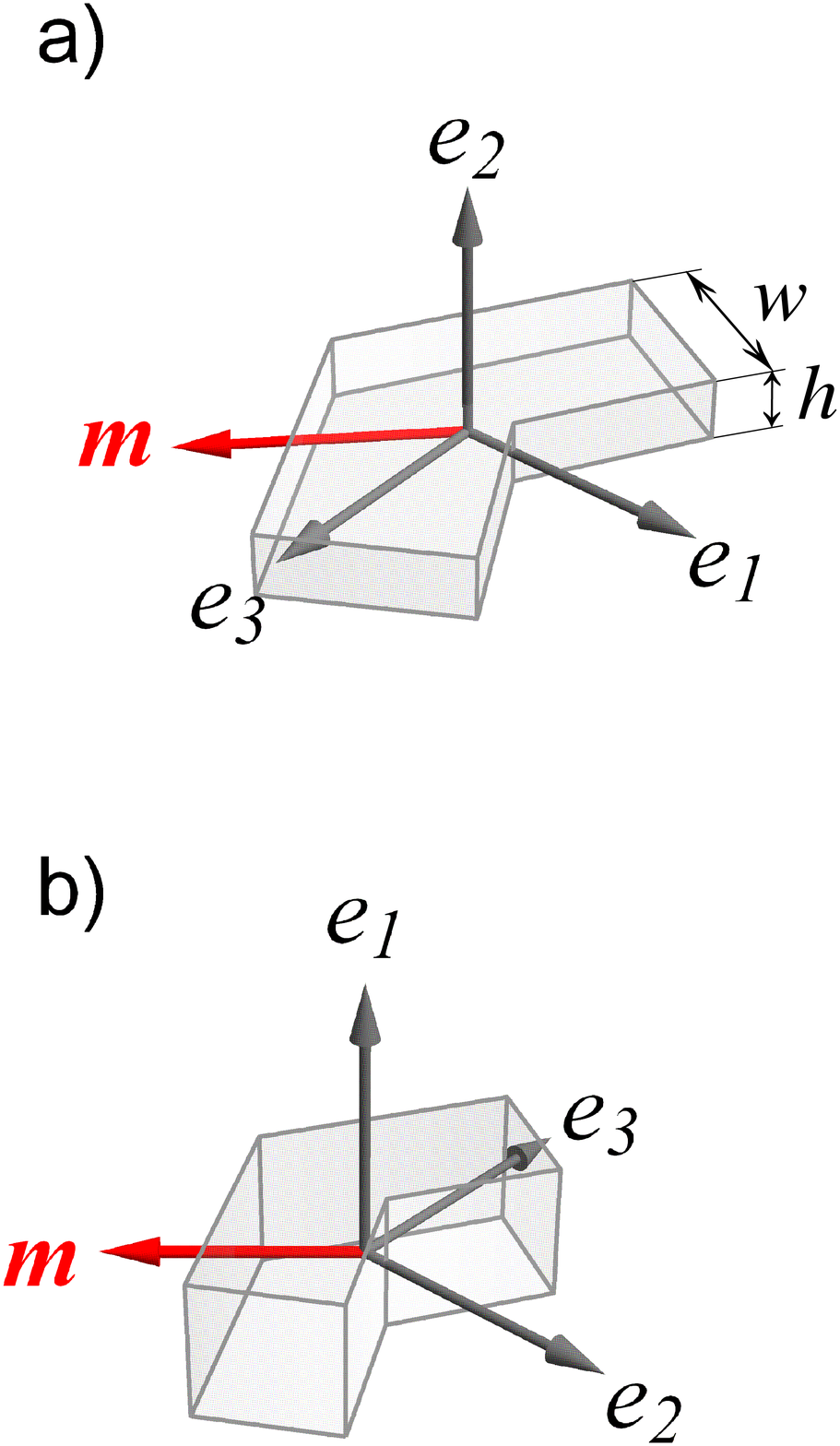}
\caption{ Planar V-shaped structures with 120-degree central angle together with their principal rotation axes $\{\bm e_1, \bm e_2, \bm e_3\}$: a) \emph{slim} structure with rectangular cross section (height-to-width aspect ratio $h$:$w$=1:3); b) \emph{chubby} structure (square cross section). The structures are magnetized in-plane and the red arrow stands for the magnetic dipolar moment $\bm m$. \label{fig:schematic}}
\end{figure}

Applying Eq.~(\ref{eq:Uz}) to the symmetric V-shaped object (see Fig.~\ref{fig:schematic}) whereas $\bCh$ has a pair of identical nonzero off-diagonal entries is straightforward. For such structures the easy rotation axis $\bm e_3$ (corresponding to the largest eigenvalue $\mathcal{F}_3$) is always parallel to the line connecting the arms of the V-shape, one minor axes coincides with the symmetry axis and another is perpendicular to the plane of the V-shape. For definiteness we choose the in-plane minor axis (along the symmetry axis) along the line bisecting the acute/obtuse angle formed by the arms of V-shape and pointing away from the vertex, as shown in Fig.~\ref{fig:schematic}.
It should be noticed however that our convention of fixing the body frame in a way that ${\mathcal F}_{1}\le{\mathcal F}_{2}\le{\mathcal F}_{3}$ can result
in the interchange of minor axes, $\bm e_1 \leftrightarrow \bm e_2$ (and $\bm e_3 \leftrightarrow -\bm e_3$ to keep the frame right-handed) upon varying the V-shape opening angle or the aspect ratio $h$:$w$ as shown in Fig.~\ref{fig:schematic}.
Considering for definiteness the case of \emph{chubby} V-shape shown in Fig.~\ref{fig:schematic}b where the only nonzero elements of $\bcalG$ are $\mathcal{G}_{13}=\mathcal{G}_{31}$, then (\ref{eq:Uz}) reduces to
\be
\frac{U_z}{\omega \ell}= \widetilde{\mathrm {Ch}}\, s_{\psi} s_{2\theta}\,, \label{eq:U1}
\ee
where $\widetilde{\mathrm {Ch}}=\mathrm{Ch}_{13}=\mathrm {Ch}_{31}=\mathcal{G}_{13}({{\cal F}_{1}}^{-1}+{{\cal F}_{3}}^{-1})/2\ell$ is the pseudo-chiral coefficient \cite{prf17,pre18}.

When the same V-shaped propeller is not turning in-sync with the field, the propulsion velocity is given by (see Appendix \ref{A}):
\be
\frac{U_z}{\ell}=\widetilde{\mathrm {Ch}}s_{\psi}s_{2\theta}\dot{\varphi}+
{\mathcal G}_{13}\left(\frac {1}{{\mathcal F}_{3}}s_{\psi}s_{\theta}\dot{\psi}+
\frac {1}{{\mathcal F}_{1}}c_{\psi}c_{\theta}\dot{\theta}\right)\,. \label{eq:U1a}
\ee
Clearly, for in-sync actuation $\dot{\theta}=\dot{\psi}=0$, $\dot{\varphi}=\omega$ and (\ref{eq:U1a}) reduces to (\ref{eq:U1}).
When the minor axes interchange (see Fig.~\ref{fig:schematic}a) the similar equations apply with $\psi\rightarrow \pi/2-\psi$, $\theta \rightarrow \pi-\theta$ and $\mathcal{G}_{23}$ replacing $\mathcal{G}_{13}$ everywhere.

\section{Hydrodynamic mobilities of planar V-structures}

\begin{figure}[tb] \centering
\includegraphics[scale=0.8]{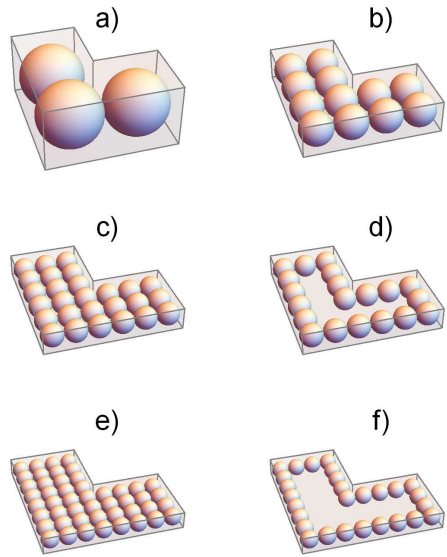}
\vskip-0.2cm
\caption{\label{fig:flats} Planar V-shapes with 90-degree central angle and varying thickness (width-to-height aspect ration, $w$:$h$). Bead-made structures approximate hydrodynamic mobilities of the planar V-shapes. a) $h$:$w$=1:1; b) $h$:$w$=1:2; c) $h$:$w$=1:3; d) $h$:$w$=1:3 (hollow structure); e) $h$:$w$=1:4; f) $h$:$w$=1:4 (hollow structure).}
\end{figure}

We apply particle-based method \cite{Filippov} for computing mobility tensors $\bcalF$, $\bcalG$, the resulting anisotropy parameters $p$, $\varepsilon$ and the chirality matrix $\bCh$ for planar V-shaped propellers. This technique is based on multipole expansion of the Lamb's spherical harmonic solution of the Stokes equations. The object is approximated by (i) monolayer of $N$ touching rigid spheres of radius $a$, as shown in Figs.~\ref{fig:flats}a--c, e or (ii) a hollow structure with beads retracing the perimeter (see Figs.~\ref{fig:flats}d,f). Thickness of the propeller is controlled by varying the number of beads approximating the object. 
This particle-based approach was previously applied for modeling self-locomotion of an undulating flexible filament \cite{BKSL13}, magnetically driven propulsion of rigid helical \cite{ML14a,Walker} and arc-shaped \cite{prf17,pre18} structures and random fractal-like aggregates \cite{scirob}.

The results of the computation are collected in Table~\ref{tab:mobilities} for different values of the central angle $\gamma$ of the V-shape.

For straight stripes ($\gamma=\pi$) 
rotation-translation coupling vanishes, $\bcalG=0$, as can be anticipated from symmetry.
Under the convention for the body frame selection (see Sec.~\ref{sec:general}), for generic value of $\gamma$ the coupling matrix $\bcalG$ has exactly two nontrivial off-diagonal elements $\mathcal{G}_{i3}=\mathcal{G}_{3i}$ with either $i=1$ or, respectively, $i=2$, depending on geometry of the structure.
We find $\mathcal{G}_{13}<0$ for V-shapes with the opening angle $\gamma=\pi/2$, while for V-shapes with the opening angle $\gamma=2\pi/3$ upon increasing the slenderness the minor axes interchange, $e_2 \leftrightarrow \bm e_1$ (as illustrated in Fig.~\ref{fig:schematic}) leading to a sudden change of sign of $\mathcal{G}_{i3}$ and $\widetilde{\mathrm {Ch}}$ in Table~\ref{tab:mobilities}, such that  $\mathcal{G}_{23}>0$ becomes the only nontrivial entry.

As one may expect for Stokes flows, the computed rotational and coupling mobilities of hollow 2D structures (in Fig.~\ref{fig:flats}d,f) are quite close to these found for the respective densely packed structures (in Fig.~\ref{fig:flats}c,e).
\begingroup
\squeezetable
\begin{table}
\begin{tabular}{|c|c|c|c|c|c|c|c|c|c|}
\hline
$\gamma$ & $h$:$w$ & f/h 
& $\widetilde{\mathcal{F}}_1$ & $\widetilde{\mathcal{F}}_2$ & $\widetilde{\mathcal{F}}_3$ & $\widetilde{\mathcal{G}}_{i3}(\times10^2)$ & $\widetilde{\mathrm{Ch}}(\times 10^2)$
& $\varepsilon$ & $p$ \\
\hline
 & 1:1 & f 
& $0.659$ & $0.756$ & $1.188$ & $-0.470$ & $-0.554$
& $0.068$ & $1.685$ \\
\cline{2-10}
 & 1:2 & f 
& $0.913$ & $1.050$ & $1.175$ & $-0.578$ & $-0.482$
& $0.070$ & $1.79$ \\
\cline{2-10}
$\pi/2$ & 1:3 & f 
& $1.046$ & $1.161$ & $1.959$ & $-0.620$ & $-0.455$
& $0.052$ & $1.78$ \\
\cline{2-10}
 & 1:3 & h 
& $1.049$ & $1.171$ & $1.971$ & $-0.607$ & $-0.443$
& $0.055$ & $1.78$ \\
\cline{2-10}
 & 1:4 & f 
& $1.131$ & $1.219$ & $2.068$ & $-0.641$ & $-0.439$
& $0.037$ & $1.76$ \\
\cline{2-10}
 & 1:4 & h 
& $1.141$ & $1.249$ & $2.105$ & $-0.633$ & $-0.428$
& $0.045$ & $1.76$ \\
\hline
\hline
 & 1:1 & f 
& $0.820$ & $0.861$ & $2.027$ & $-0.933$ & $-0.799$
& $0.024$ & $2.41$ \\
\cline{2-10}
 & $\;\;$1:2\footnote{this sample is used in calculations throughout the paper} & f 
& $1.069$ & $1.078$ & $2.945$ & $-1.058$ & $-0.675$
& $0.004$ & $2.74$ \\
\cline{2-10}
 & 1:3 & f 
& $1.164$ & $1.206$ & $3.326$ & $1.097$ & $0.620$
& $0.017$ & $2.81$ \\
\cline{2-10}
$2\pi/3$ & 1:3 & h 
& $1.173$ & $1.209$ & $3.340$ & $1.110$ & $0.620$
& $0.015$ & $2.80$ \\
\cline{2-10}
 & 1:4 & f 
& $1.275$ & $1.376$ & $3.994$ & $1.438$ & $0.702$
& $0.038$ & $3.02$ \\
\cline{2-10}
 & 1:4 & h 
& $1.308$ & $1.388$ & $4.041$ & $1.513$ & $0.732$
& $0.030$ & $3.00$ \\
\hline
\hline
 & 1:1 & f 
& $0.942$ & $0.942$ & $3.317$ & $0$ & $0$
& $0$ & $3.52$ \\
\cline{2-10}
 & 1:2 & f 
& $1.037$ & $1.088$ & $4.553$ & $0$ & $0$
& $0.024$ & $4.29$ \\
\cline{2-10}
$\;\;\pi$\footnote{rectangular stripes} & 1:3 & f 
& $1.086$ & $1.195$ & $5.048$ & $0$ & $0$
& $0.048$ & $4.43$ \\
\cline{2-10}
 & 1:3 & h 
& $1.095$ & $1.199$ & $5.057$ & $0$ & $0$
& $0.045$ & $4.42$ \\
\cline{2-10}
 & 1:4 & f 
& $1.113$ & $1.268$ & $5.298$ & $0$ & $0$
& $0.065$ & $4.47$ \\
\cline{2-10}
 & 1:4 & h 
& $1.142$ & $1.278$ & $5.338$ & $0$ & $0$
& $0.056$ & $4.43$ \\
\hline
\end{tabular}
 \caption{Comparison of hydrodynamic mobilities of planar V-shaped structures.
 $\gamma$ is the opening angle; `f' and `h' stand for either filled or hollow cluster, respectively;
$h$:$w$ is the height-to-width aspect ratio; $\widetilde{\mathcal{F}}_{i}$ are the
eigenvalues of the respective dimensionless rotational mobility tensor 
$\eta \ell^3 \bcalF$; $\widetilde{\mathcal{G}}_{i3}$ ($i=1$ or $2$) are the unique nonzero off-diagonal elements of the symmetrized coupling matrix
$\eta\ell^2 \bcalG$; $\widetilde{\mathrm {Ch}}=\mathcal{G}_{i3}({{\cal F}_{i}}^{-1}+{{\cal F}_{3}}^{-1})/2\ell$ is the pseudo-chirality coefficient; $\varepsilon$ and $p$ are the transversal and longitudinal rotational anisotropy parameters, respectively.
\label{tab:mobilities}}
\end{table}
\endgroup

\section{Actuation of V-shaped propellers by a conical magnetic field}
In this section we shall consider a number of analytically tractable cases and approximate solutions. In the analysis below we assume, for definiteness, the orientation of the principal axes as shown in Fig.~\ref{fig:schematic}b. Modification of the present analysis for the slim structures (as in Fig.~\ref{fig:schematic}a) is straightforward.

\subsection{Magnetization along the symmetry $\bm e_2$-axis \label{sec:e2}}
Assuming magnetization along the symmetry axis $\bm e_2$, i.e., $n_1=n_3=0$ and $n_2=n_\perp=1$, we have $\Phi=\alpha=\pi/2$. Then from (\ref{eq:2b}) it follows that $\psi=\pm \pi/2$ (it can be readily seen that $\sin{\theta}=0$ is not a solution). Substituting these values of the magnetization angles and $\psi$ into Eqs.~(\ref{eq:1b},{\ref{eq:3b}}) we obtain
\be
s_{2\theta}=\pm \frac{2\delta}{\widetilde{\omega} \left(1+\varepsilon-p^{-1}\right)}\,,\label{eq:sol}
\ee
which imposes restriction on the value of $\delta$ (or $\omega$) for which in-sync solution materializes. The $\pm$ signs correspond to two rotational solutions with acute and obtuse wobbling angle $\theta$, respectively.
Then, knowing $\theta$ the Euler angle $\widehat{\varphi}$ can be found from (\ref{eq:1b}) as
\be
s_{\widehat{\varphi}}=\pm (1+\varepsilon) {\widetilde \omega}-\delta \cot{\theta}\;. \label{eq:varphi}
\ee

Notice that as $\bm m$ is aligned with $\bm e_2$, the two (dual) rotational states in Eqs.~(\ref{eq:sol}--\ref{eq:varphi}) are equivalent and cannot be distinguished due to arbitrariness in the choice of orientation of the principal axes: rotating the V-shape by $\pi$ around $\bm e_2$ brings the object to itself.
Thus, one may consider only the solution corresponding to an acute wobbling angle $\theta<\pi/2$. The positive value of $s_{2\theta}$ corresponds to the two distinct values of the precession angle $\theta$, while only one (the smaller of the two, $\theta<\pi/4$) proves to be stable (see the paragraph on stability below).

Clearly, the in-sync solution Eqs.~(\ref{eq:sol}--\ref{eq:varphi}) persists in a limited range of actuation frequencies, $\widetilde{\omega}_*<\widetilde{\omega}<\widetilde{\omega}_\mathrm{s\mbox{-}o}$, where
\be
\widetilde{\omega}_*=\frac{2\delta}{1+\varepsilon - p^{-1}}\,,\label{eq:omega_cr}
\ee
is imposed by Eq.~(\ref{eq:sol}) at $\theta=\pi/4$ and the step-out frequency, $\widetilde{\omega}_\mathrm{s\mbox{-}o}$, by
(\ref{eq:varphi}) at $\widehat{\varphi}_\mathrm{s\mbox{-}o}=\pi/2$.
The step-out frequency can in turn be determined as follows. Using the identity $\sin{2\theta}=2\cot\theta/(1+\cot^2\theta)$ and substituting
$\cot{\theta}$ from Eq.~(\ref{eq:varphi}) for $\sin{\widehat{\varphi}}=1$ into (\ref{eq:sol}), we obtain the quadratic equation for $\widetilde{\omega}_\mathrm{s\mbox{-}o}$:
\[
(1+\varepsilon)\widetilde{\omega}_\mathrm{s\mbox{-}o}^2-(1+p+p\varepsilon)\widetilde{\omega}_\mathrm{s\mbox{-}o}+p(1+\delta^2)=0\:.
\]
The larger root gives the step-out frequency:
\be
\widetilde{\omega}_\mathrm{s\mbox{-}o}=\frac{1+\epsilon_p+\sqrt{(1+\epsilon_p)^2-4 (1+\delta^2) \epsilon_p}}{2 (1+\varepsilon)}\,, \label{eq:omega_so}
\ee
where $\epsilon_p=p(1+\varepsilon)$. With $\widetilde{\omega}_\mathrm{s\mbox{-}o}$ at hand, one can readily determine the precession angle at the step-out, $\theta_\mathrm{s\mbox{-}o}$, from, e.g., Eq.~\ref{eq:sol}.
It can also be shown that at the step-out the angle $\beta$ between magnetization $\bm m$ and the field $\bm H$ attains its maximal value of $\pi/2$, maximizing the magnetic torque.

\begin{figure} \centering
\includegraphics[scale=0.8]{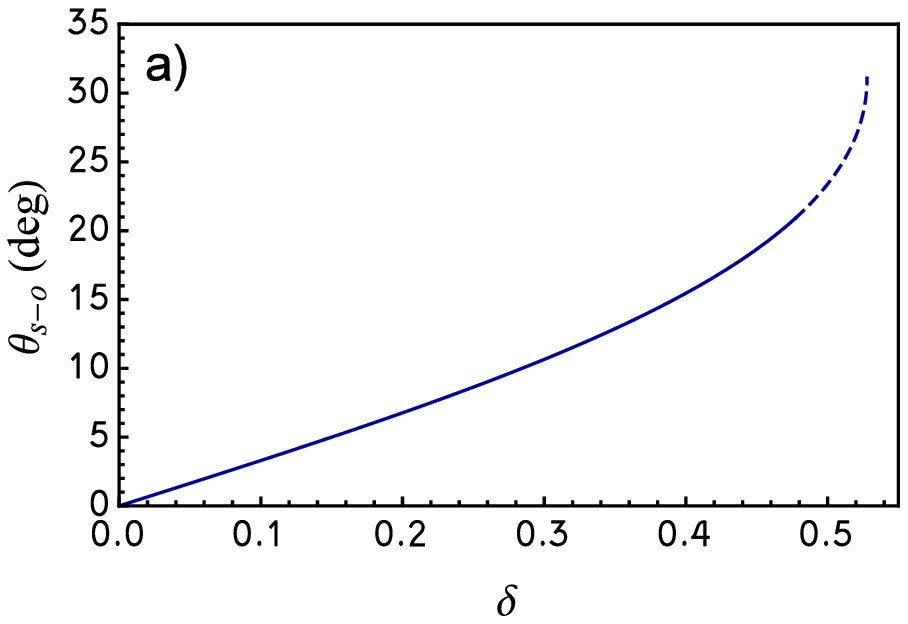}\\
\vskip2mm
\includegraphics[scale=0.8]{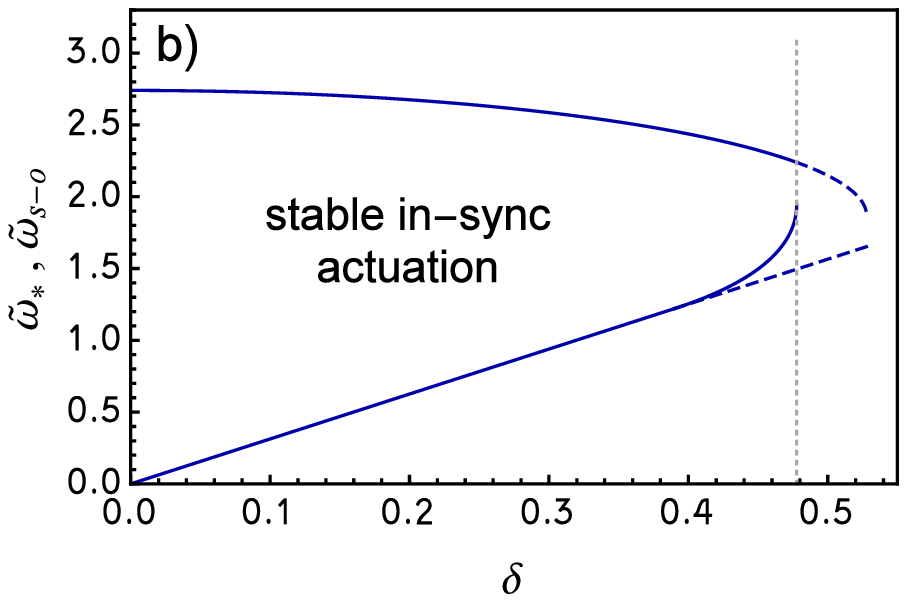}
\caption{In-sync actuation of the V-shape with the central angle $\gamma=120^\circ$ and cross-section aspect ratio $h$:$w$=1:2 magnetized along $\bm -e_2$. (a) Wobbling angle at the step-out $\theta_\mathrm{s\mbox{-}o}$ vs. $\delta$. (b) The dimensionless transition frequency, $\omega_*/\omega_0$ (lower curve), and the step-out frequency, $\omega_\mathrm{s\mbox{-}o}/\omega_0$ (upper curve) vs. $\delta$. Stable in-sync solutions materialize in a limited range of actuation frequency (between the two \emph{solid} lines), no stable in-sync solutions can be found for $\delta\gtrsim 0.478$ (vertical dotted line); dashed segments mark \emph{unstable} in-sync solutions. \label{fig:stepout} }
\end{figure}

As an example, for the V-structure with a cross-section aspect ratio $h$:$w$=1:2 and central angle  $\gamma=120^\circ$ (see Table~\ref{tab:mobilities}), we have $p=2.74$ and $\varepsilon=0.004$. For this sample propeller, the wobbling angle at the step-out, $\theta_\mathrm{s\mbox{-}o}$, increases with $\delta$ up to a maximum value $\approx 31.1^\circ$ at $\delta=0.53$ (see Fig.~\ref{fig:stepout}a); above this value of $\delta$ no in-sync solution exist. The step-out frequency in Eq.~(\ref{eq:omega_so}) slightly diminishes with $\delta$ (see Fig.~\ref{fig:stepout}b). For instance for $\delta=0.1$ and $0.4$ we have ${\widetilde\omega}_\mathrm{s\mbox{-}o}=2.72$ and $2.44$, respectively. Notice that no \emph{stable} in-sync solutions can be found at $\delta \gtrsim 0.478$ (see Fig.~\ref{fig:stepout}b and the stability analysis below).

Substituting the steady-state solution for $s_{2\theta}$ from (\ref{eq:sol}) and $\psi=\pm \pi/2$ into (\ref{eq:U1}) we readily obtain the in-sync propulsion velocity of a magnetic V-shape:
\be
\frac{U_z}{\omega_0 \ell}=\widetilde{\mathrm {Ch}} \frac{2\delta}{\left(1+\varepsilon - p^{-1}\right)}\,,\label{eq:U2}
\ee
which surprisingly is independent of the actuation frequency.

In contrast to helical propellers which swim the best when precession is minimized (similar to a corkscrew twirling around its long axis), efficient propulsion of planar structures requires considerable precession \cite{prf17}. Since the wobbling angle, $\theta$, diminishes with the actuation frequency as $s_{2\theta}\sim 1/\omega$, while the propulsion velocity $U_z\sim \omega s_{2\theta}$, reduction of the precession angle is compensated exactly by the increasing rotation rate, rendering $U_z$ constant in a limited range of in-sync actuation frequencies $\omega_*<\omega<\omega_\mathrm{s\mbox{-}o}$. We consider Eq.~(\ref{eq:U2}) as a major result of the present paper.  Animations of the in-sync driven rotation and propulsion of the sample propeller (with $\gamma=120^\circ$ and $h$:$w$=1:2, see Table~\ref{tab:mobilities}) are provided in \cite{SM} for $\delta=0.3$ and several value of the actuation frequency $\omega/\omega_0$. Notice that in the movies the magnetic moment $\bm m$ rotates in the $xy$-plane of the field. A simple way to show that, is to note that for in-sync solution we have $\bm{\mathit{\Omega}}$~$\|$~$\hat{\bm z}$, while the magnetic torque $\bm L=\bcalF^{-1} \cdot \bm{\mathit{\Omega}}\perp\bm m$. Thus it follows that $\bm m \cdot \bcalF^{-1} \cdot \hat{\bm z}=0$ and since $\bcalF$ is symmetric it means that $\bcalF^{-1}\cdot\bm m$~$\perp$~$\hat{\bm z}$. For an object magnetized along one of the principal axes (i.e., eigenvectors of $\bcalF$) we have $\bcalF^{-1}\cdot \bm m$~$\|$~$\bm m$ so that $\bm m$~$\perp$~$\hat{\bm z}$.

\begin{figure}[h] \centering
\includegraphics[scale=0.9]{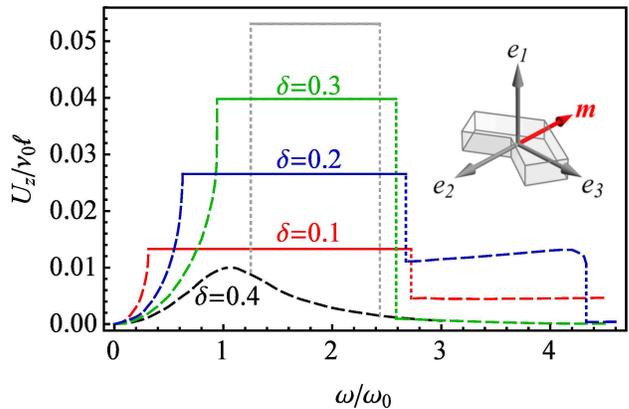}
\caption{The dimensionless propulsion velocity, $U_z/\nu_0\ell$, of the V-shape with the central angle $\gamma=120^\circ$ and cross-section aspect ratio $h$:$w$=1:2 magnetized along $\bm -e_2$ (see the inset), as a function of the scaled actuation frequency, $\omega/\omega_0$, for different magnitudes of the axial magnetic field ($\delta$). Solid lines stand for the constant in-sync solutions (\ref{eq:U2}) and long-dashed lines to asynchronous solutions.}
\label{fig:U}
\end{figure}

The frequency dependence of the dimensionless propulsion velocity, $U_z/\nu_0\ell$ \cite{note0}, of the same sample V-shape, where $\nu_0=\omega_0/2\pi$ stands for the characteristic (cyclic) frequency, is depicted in Fig.~\ref{fig:U} for several values of $\delta$. The Euler angles were obtained by numerical integration of Eqs.~(\ref{eq:1})-(\ref{eq:3}) and the propulsion velocity computed using Eq.~(\ref{eq:U1a}).
In agreement with the theory, the numerical solution produces constant propulsion velocity (\ref{eq:U2}) in a finite range of in-sync actuation frequencies in accordance with Fig.~\ref{fig:stepout}b. At higher values of $\delta=0.4$ the basin of attraction of the constant in-sync solution (solid gray line) becomes narrow and for random initial orientation the asynchronous solution materializes for all frequencies (dashed black curve in Fig.~\ref{fig:U}). At low actuation frequency $\omega<\omega_*$ the solutions (dashed lines) are quasi-synchronous, meaning that Euler angles oscillate periodically about some constant values, so that on average the propeller rotates
in-sync with the actuation frequency and its propulsion velocity is insensitive to the initial orientation.
Above the step-out frequency the V-shape cannot catch up with fast rotating magnetic field and, as a result, it does not turn in-sync with the field. In this regime we found that the solution converges to a stable closed orbit solution (limit cycle) in $(\theta, \psi)$-plane (with $\widehat\varphi$ periodic modulo $2\pi$) with average propulsion velocity described by the dashed line \cite{note3}. It should be noted however that the velocity  oscillates strongly over the period of this limit cycle.
Moreover it turns out that the convergence to the limit cycle is quite slow so that the standard deviation from the mean velocity for $\delta=0.2$ at $\omega/\omega_0=3$ is $\sim 10$\% when averaging over $100$ field revolutions while it drops to $\sim 1$\% when averaging over $1000$ periods.

Notice that reversing the magnetization, $\bm m \rightarrow -\bm m$ yields a parity-transformed object having the reverse propulsion
velocity, $U_z \rightarrow -U_z$.  It can also be shown explicitly from Eqs.~(\ref{eq:1b})-(\ref{eq:3b}) since taking $\alpha=-\pi/2$ and $n_2=-1$ yields,
as before,  $\psi=\pm \pi/2$, while the corresponding $\pm$ signs in Eqs.~(\ref{eq:sol}) and in (\ref{eq:varphi}) change to $\mp$.
Thus, the propulsion velocity $U_z\sim s_{2\theta} s_\psi$ in (\ref{eq:U1}) changes sign.

It should also be stressed that net propulsion occurs for the less symmetric ($\hat{P}$-)chiral propeller \cite{pre18}, while ($\hat{P}$-)achiral propeller
(i.e., $\bm m$ oriented along $\bm e_3$ or $\bm e_1$) yields no propulsion even when a constant $H_z$ field is present, as we shall see below. Please recall that for $\delta=0$ magnetization along any principal rotation axis yielded no net propulsion \cite{pre18}, while here we have shown that
adding a static field along the field-rotation axis results in unidirectional propulsion similar
to magnetic helices.\\

\noindent\emph{Stability of the in-sync solutions}. To study stability of the above in-sync solution we substitute  $n_1=n_3=0$, $n_2=n_\perp=1$ and $\Phi=\alpha=\pi/2$ into
the Eqs.~({\ref{eq:1})-(\ref{eq:3}}) governing the rotational dynamics and obtain:
\ba
&&\textstyle\frac{1+\varepsilon}{\omega_0}(\dot{\varphi}s_{\theta}s_{\psi}+\dot{\theta}c_{\psi})=s_{\theta}s_{\widehat{\varphi}} + \delta c_\theta\,,   \label{eq:1c} \\
&&\textstyle\frac{1-\varepsilon}{\omega_0}(\dot{\varphi}s_{\theta}c_{\psi}-\dot{\theta}s_{\psi})=0\,,    \label{eq:2c} \\
&&\textstyle\frac{1}{p\omega_0}(\dot{\varphi}c_{\theta}+\dot{\psi})=-c_{\psi} c_{\widehat{\varphi}}  + s_{\psi}(s_{\widehat{\varphi}} c_{\theta}-\delta s_\theta) \,.  \label{eq:3c}
\ea
We further perturb the steady solution by adding small disturbances to the steady-state values of the angles:
\[
\theta=\theta_0+\theta_1 \re^{\lambda t},\; \varphi={\widehat{\varphi}}_0+\omega t+\varphi_1 \re^{\lambda t},\; \psi=\frac{\pi}{2}+ \psi_1 \re^{\lambda t}\,,
\]
where $\theta_0<\pi/2$ and $\widehat{\varphi}_0$ are the steady in-sync solutions of Eqs.~(\ref{eq:sol}) and (\ref{eq:varphi}), respectively, and $\lambda$ is the perturbation growth rate. Substituting this ansatz into (\ref{eq:1c}-\ref{eq:3c}) and linearizing over the the perturbation amplitudes, $\bm u=(\theta_1, \psi_1, \varphi_1)$, we readily obtain the homogeneous system of equations, $P \bm u=0$, where
\ba
P=\begin{pmatrix} \delta \csc{\theta_0} & 0 & s_{\theta_0} (\widetilde{\lambda} +\widetilde{\lambda} \varepsilon -c_{\widehat{\varphi}_0})  \\ \widetilde{\lambda} & \widetilde{\omega} s_{\theta_0} & 0 \\
p \delta \sec{\theta_0} & \widetilde{\lambda}-p c_{\widehat{\varphi}_0} & c_{\theta_0}  (\widetilde{\lambda}-p c_{\widehat{\varphi}_0})\end{pmatrix},\nonumber
\ea
where $\widetilde{\lambda}=\lambda/\omega_0$.
The solvability condition $\det P=0$ yields the cubic equation $\widetilde{\lambda}^3+q \widetilde{\lambda} ^2+r\widetilde{\lambda} +s=0$ where the coefficients after some algebra reduce to
\ba
q&=&-\frac{(1+p+\varepsilon p) \cos{\widehat{\varphi}_0}}{1+\varepsilon}\,, \nonumber \\
r&=&\frac{p \cos^2{\widehat{\varphi}_0}+\delta \widetilde{\omega}  (\cot{\theta_0}-p \tan{\theta_0} (1+\varepsilon))}{1+\varepsilon}\,, \nonumber \\
s&=&-\frac{2 p \delta \widetilde{\omega} \cot{2 \theta_0} \cos{\widehat{\varphi}_0}}{1+\varepsilon}\,. \nonumber
\ea
In general, the stability criterion, i.e., real part of all three roots $\widetilde{\lambda}$ should be negative, requires $q, r, s>0$ together with the condition $qr>s$ \cite{note2}.

The condition $q>0$ readily gives $\cos{\widehat{\varphi}_0}<0$. This condition is also evident from minimizing the magnetic energy, which in this case reduces to $E=-\bm m\cdot \bm H=mH c_{\widehat{\varphi}_0}$. Notice that this condition ceases to hold exactly at the frequency $\omega_\mathrm{s\mbox{-}o}$.
The condition $s>0$ then yields  $\cot{2\theta_0}> 0$.
Since we consider the solution with $\sin{2\theta_0}>0$, it follows that $\cos{2\theta_0}>0$ resulting in $0<\theta_0<\pi/4$.
Introducing $x\equiv \cot\theta_0>1$ and using (\ref{eq:sol}) we can write
\[
s_{2\theta_0} = \frac{2x}{1+x^2}=\frac{2\delta p}{\widetilde{\omega}(p+p\varepsilon-1)} \equiv A^{-1}\,.
\]
Substituting $x = A + \sqrt{A^2-1} > 1$ into the expression for $r (1+\varepsilon) = c_{\widehat{\varphi}_0}^2 +\widetilde{\omega} \delta (x/p-(1+\varepsilon)/x)$ and using  $s_{\widehat{\varphi}_0} = (1+\varepsilon)\widetilde{\omega}-\delta x$ to eliminate $\widehat{\varphi}_0$, we obtain a condition on $x$ from $r>s/q>0$. The final stability criterion on $\theta_0$ reads
\be
\cot{\theta_0}> \max\left[1,\; \frac{2 p \delta \tilde\omega (1+\epsilon_p+\epsilon_p^2)}{p^2(1+\delta^2)(1+\epsilon_p)-2\tilde\omega^2}\right]\,. \label{eq:stability}
\ee
where $\epsilon_p=p(1+\varepsilon)$.
The second condition in (\ref{eq:stability}) imposes stricter restrictions on $\theta_0$ and $\omega$ in comparison to those imposed by Eqs.~(\ref{eq:sol}), (\ref{eq:varphi}) at higher values of $\delta$ and, in fact, no \emph{stable} in-sync solutions can be found already for $\delta\gtrsim 0.478$ (see Fig.~\ref{fig:stepout}).

\begin{figure}[htb]
\centering
\vskip2mm
\includegraphics[scale=0.8]{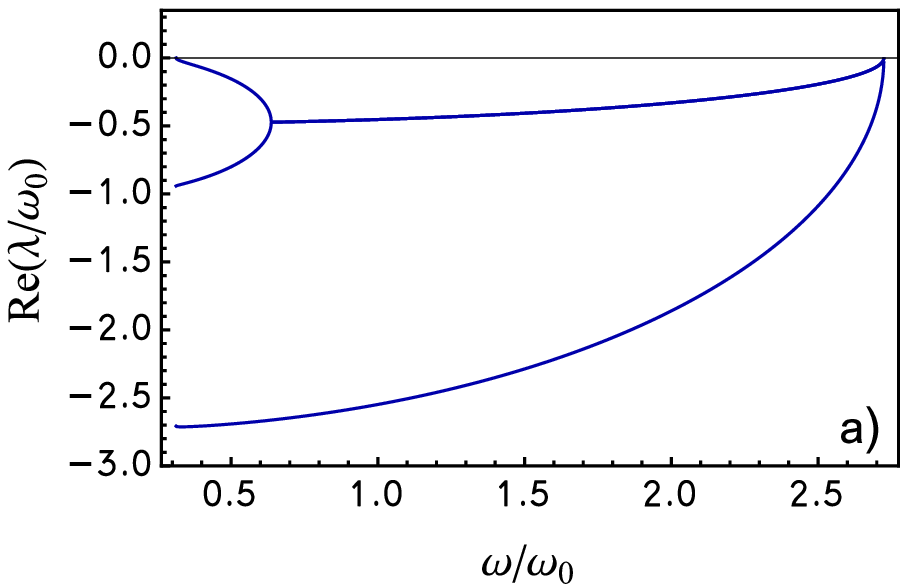}\\
\vskip2mm
\includegraphics[scale=0.8]{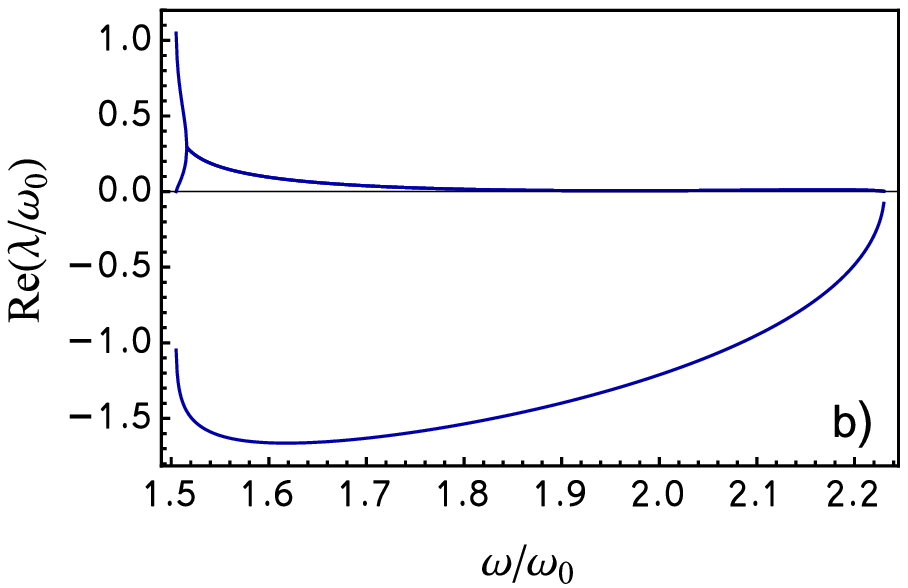}
\caption{\label{fig:lambda} Real part of the perturbation growth rate, $\mathrm{Re}(\lambda/\omega_0)$, vs.
the actuation frequency $\omega/\omega_0$ for chubby the V-structure in Fig.~\ref{fig:U}: a) $\delta=0.1$ (stable); b) $\delta=0.48$ (unstable).}
\end{figure}

The real part of the dimensionless growth rates  $\mathrm{Re}(\lambda/\omega_0)$ corresponding to the in-sync solution with $\theta_0<\pi/4$ for $\delta=0.1$ (stable)
and $\delta=0.48$ (unstable) are depicted in Figs.~\ref{fig:lambda}a,b vs. scaled frequency $\omega/\omega_0$ for the V-shaped propeller shown in the inset in Fig.~\ref{fig:U}.

\subsection{Magnetization along the rotation easy-axis $\bm e_3$}
When $\bm m$ is oriented along $\bm e_3$, i.e., $n_3=1$ and $n_1=n_2=n_\perp=0$, we have $\Phi=0$. Then from Eq.~(\ref{eq:3b}) we immediately have $c_\theta=0$ and $\theta=\pm \pi/2$. then from (\ref{eq:U1}) it follows that $U_z=0$. This regime corresponds to \emph{tumbling}. \\

\subsection{Off-plane magnetization along $\bm e_1$-axis}
When $\bm m$ is  parallel to $\bm e_1$, i.e., $n_1=n_\perp=1$ and $n_2=n_3=0$, we have $\Phi=\pi/2$ and $\alpha=0$. Then from Eq.~(\ref{eq:1b}) we find that $(1+\varepsilon) \widetilde{\omega} s_\theta s_\psi=0$, i.e., $\psi=0$ or $\pi$. For these values of $\psi$ we have $U_z=0$ in (\ref{eq:U1}). Notice that the wobbling angle $\theta$ found from (\ref{eq:2b}) and (\ref{eq:3b}) is nontrivial:
\be
s_{2\theta}=\mp \frac{2\delta}{\widetilde{\omega} \left(1-\varepsilon-p^{-1}\right)}\,,\label{eq:sol1}
\ee
where $\mp$ holds for $\psi=0$ and $\pi$, respectively. This means that the object will undergo precession with a finite angle $\theta$ which is not accompanied by net propulsion in contrast to magnetization along the symmetry axis \cite{note1}.

\subsection{Magnetization in ${\bm e}_1 {\bm e}_2$-plane, approximate solution}
When $\bm m$ is oriented perpendicular to the rotation easy-axis $\bm e_3$ (i.e. $n_3=0$, $\Phi=\pi/2$), it is possible to find an approximate solution assuming $\varepsilon=0$ (i.e. cylindrical approximation). In this case it readily follows from Eqs.~(\ref{eq:1b})--(\ref{eq:2b}) that $t_\psi=-t_\alpha$, meaning $\psi=-\alpha$ or $\psi=\pi-\alpha$. Then subtracting Eq.~(\ref{eq:3b}) multiplied by $s_\theta$ from (\ref{eq:2b}) multiplied by $c_\theta/c_\psi$ we obtain
\be
s_{2\theta}=\pm \frac{2\delta}{\widetilde{\omega} \left(1-p^{-1}\right)}\,,\label{eq:theta1}
\ee
where $\pm$ corresponds respectively to $\psi=\pi-\alpha$ (for $0<\theta<\pi/2$ ) and $\psi=-\alpha$ (for $\pi/2<\theta<\pi$).
Thus using $s_\psi=\pm s_\alpha$ and (\ref{eq:U1}) at the zeroth order approximation in $\varepsilon$ we arrive at
\be
\frac{U_z}{\omega_0 \ell} \approx \widetilde{\mathrm {Ch}} s_\alpha \frac{2\delta}{\left(1- p^{-1}\right)}\,.\label{eq:U3}
\ee
The transverse (to $\bm e_3$) magnetization also results in unidirectional propulsion with constant velocity as was found for the in-plane magnetization along $\bm e_2$ in Sec.~\ref{sec:e2}. Notice that the propulsion velocity in (\ref{eq:U3}) attains its maximum value at $\alpha=\pi/2$, corresponding to $\bm m \|\bm e_2$, which agrees with the exact result (\ref{eq:U2}) up to $\mathcal{O}(\varepsilon)$.

The in-sync propulsion (\ref{eq:U3}) persists in a limited range of actuation frequencies, $\widetilde{\omega}_*<\widetilde{\omega}<\widetilde{\omega}_\mathrm{s\mbox{-}o}$.
Considering a solution corresponding to an acute precession angle $\theta$ and  noting that the solution of the rotational problem assuming cylindrical anisotropy does not depend on $\alpha$, we find that $\widetilde{\omega}_*$ and $\widetilde{\omega}_\mathrm{s\mbox{-}o}$ are given to the first approximation by the respective expressions (\ref{eq:omega_cr}) and (\ref{eq:omega_so}) at $\varepsilon=0$, such that
\be
\frac{2\delta}{1-p^{-1}} < \widetilde{\omega} <\frac{1}{2} (1+p+\sqrt{(p-1)^2-4p \delta^2})\,.\label{eq:omega_so2}
\ee

\subsection{Arbitrary magnetization, $\mathcal{O}(\delta)$ asymptotic theory}

Although we could not find a closed form solution for an arbitrary magnetization for finite $\delta$, it is possible to to take advantage of the cylindrical approximation  ($\varepsilon\simeq 0$) for which an exact analytical solution is known for an in-plane rotating field for $\delta=0$ (see \cite{prf17}), and construct a small $\mathcal{O}(\delta)$ expansion around it. This approximation provides closed-form solutions for the low-frequency ``tumbling" ($\theta=\pi/2$) and high-frequency ``wobbling" ($\theta< \pi/2$) regimes of in-sync actuation. The explicit form of the ``tumbling" solution at low frequencies, $0<\widetilde{\omega}<\widetilde{\omega}_\mathrm{t\mbox{-}w}$, where $\widetilde{\omega}_\mathrm{t\mbox{-}w}=c_{\Phi}$, is given by
\be
\theta=\pi/2\,,\,\,\psi=-\alpha\,,\,\,{\widehat{\varphi}}=-\Phi+\arccos{\widetilde{\omega}}\,. \label{eq:t1}
\ee
At higher frequencies $\widetilde{\omega}_\mathrm{t\mbox{-}w}<\widetilde{\omega}< \widetilde{\omega}_\mathrm{s\mbox{-}o}$, where the step-out frequency is $\widetilde{\omega}_\mathrm{s\mbox{-}o}=\sqrt{c_{\Phi}^2+s_{\Phi}^2 p^2}$, the two symmetric modes of the wobbling solution are given by
\ba
&\theta_1=\arcsin\left({\frac{c_{\Phi}}{\widetilde{\omega}}}\right),\;\psi_1=-\alpha-\arcsin\left({\frac{c_{\theta_1}{\widetilde{\omega}}}{p\:s_\Phi}}\right), & \label{eq:w1}\\
&\theta_2=\pi -\theta_1,\,\,\,\psi_2=-2\alpha-\psi_1\,. & \label{eq:w2}
\ea
whereas $\widehat{\varphi}_1=\widehat{\varphi}_2=0$.

We therefore look for a solution to the rotational problem for finite $0<\delta\ll 1$ as regular perturbation in $\delta$ via $\theta=\theta^0+\delta \theta^1+\ldots$, $\psi=\psi^0+\delta \psi^1+\ldots$, $\widehat{\varphi}=\delta \varphi^1+\ldots$, where the superscript ``0" stands for the zero-order ``tumbling" solution (\ref{eq:t1}). Substituting these expansions into Eqs.~(\ref{eq:1b})--(\ref{eq:3b}), collecting $\mathcal{O}(\delta)$ terms and solving the resulting system of equations for the first-order corrections to the Euler angles in ``tumbling" regime gives:
\be
\{\theta^1,\, \psi^1\} = -\frac{\{p s_\Phi,\, c_\Phi\}}{(p s_\Phi s_{\widehat{\varphi}_0}-c_\Phi c_{\widehat{\varphi}_0}+\widetilde{\omega})}\,,\quad
\varphi^1=0\,.
\label{eq:asymp1}
\ee
Analogous expansion around the zero-order ``wobbling" solution in Eq.~(\ref{eq:w1})--(\ref{eq:w2}) yields the following $\mathcal{O}(\delta)$-correction (identical for both solution branches):
\be
\{\theta^1,\,\psi^1\}= \frac{\{\widetilde{\omega} \sqrt{p^2 s^2_\Phi+c^2_\Phi -\widetilde{\omega}^2},\,\, \widetilde{\omega}^3 \sec{\Phi}\}}
{(p-1)(c^2_{\Phi}-\widetilde{\omega}^2)}\,.
\label{eq:asymp2}
\ee
and
\[
\varphi^1=\frac{\left(\widetilde{\omega}^2+(p-1) c^2_\Phi\right)\sqrt{\widetilde{\omega}^2-c^2_\Phi}\sec{\Phi}}{(p-1)(c^2_{\Phi}-\widetilde{\omega}^2)}\,.
\]
Although the regular asymptotic expansions in (\ref{eq:asymp1})--(\ref{eq:asymp2}) break down in the vicinity of of the tumbling-to-wobbling transition, they are expected to closely approximate the solution elsewhere, i.e., except for the vicinity of $\widetilde{\omega}_\mathrm{t\mbox{-}w}$.

\begin{figure} [htb]\centering
\includegraphics[scale=0.95]{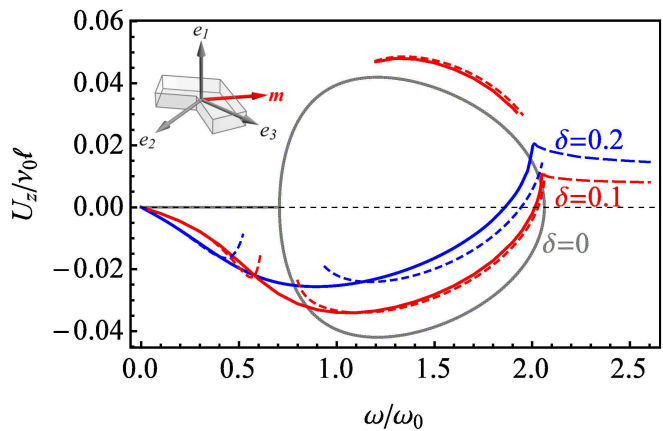} 
\caption{The dimensionless propulsion velocity of the V-shape structure (the same as in Fig.~\ref{fig:U}) with in-plane magnetization for $\Phi=\pi/4$, $\alpha=-\pi/2$ (see the inset) as a function of the actuating frequency $\omega/\omega_0$ for different magnitudes of the constant magnetic field: $\delta=0$ (gray), $\delta=0.1$ (red) and $\delta=0.2$ (blue). Solid lines stand for the stable in-sync numerical solution, long-dashed lines to asynchronous solutions and short-dashed lines to $\mathcal{O}(\delta)$ asymptotic approximation.}
\label{fig:U2}
\end{figure}

To illustrate the applicability of this approximation we compute the propulsion velocity of V-shaped propeller magnetized in its plane by numerically integrating Eqs.~(\ref{eq:1})-(\ref{eq:3}) to find the solution of the rotational problem, and then compare these numerical result with $\mathcal{O}(\delta)$ asymptotic prediction.
When the V-shape is magnetized in its plane (but not along one of the principal rotation axes), it can efficiently propel even if actuated
by an in-plane rotating magnetic field (i.e., for $\delta=0$)  above certain actuation frequency in a ``wobbling" regime by a \emph{spontaneous} symmetry breaking \cite{pre18}, whereas it can move in either ($\pm z$) direction depending on its initial orientation. The resultant symmetric velocity-frequency dependence is
depicted in Fig.~\ref{fig:U2} (gray solid line). Turning on the constant $H_z$ field removes the degeneracy between two branches of the symmetric pitchfork ``balloon" that bifurcates into a continuous (lower) branch and an isolated (upper) branch materializing over a limited range of actuation frequencies (red solid lines). In other words, symmetric pitchfork ``balloon" is structurally unstable and it bifurcates similarly to the imperfection-sensitivity diagram of compressional buckling of an elastic rod \cite{buckling}.

The frequency range of the isolated solution branch shrinks as $\delta$ increases and disappears completely for $\delta\approx 0.18$ (see the blue curve in Fig.~\ref{fig:U2}).
The vanishing of the second branch of the solution can potentially be used for enhanced passive control of propulsion of planar magnetic micro/nanomotors.
The effect of turning on the static $H_z$ field is analogous to the effect of transverse rotational anisotropy of the magnetic propeller driven by an in-plane rotating field with
$\varepsilon$ (rather than $\delta$) playing a role of the ``imperfection parameter" in bifurcation of the symmetric pitchfork ``balloon" dependence (see Fig.~4 in \cite{prf17}).

The dashed color lines in Fig.~\ref{fig:U2} correspond to small-$\delta$ approximation for the in-sync velocity (\ref{eq:U1}) upon substituting the first-order expansions for the Euler angles, $\theta=\theta^0+\delta \theta^1$, $\psi=\psi^0+\delta \psi^1$ using Eqs.~(\ref{eq:t1}) and (\ref{eq:asymp1}) at low frequencies or Eqs.~(\ref{eq:w2}) and (\ref{eq:asymp2}) at high frequencies. It can be readily seen that the agreement between the numerical results and $\mathcal{O}(\delta)$ asymptotic theory is quite accurate, except at the the vicinity of
$\widetilde{\omega}_\mathrm{t\mbox{-}w}$ where both expansions break down. The step-out frequency is only slightly altered by finite $\delta$ and $\varepsilon$ and can be predicted quite accurately by the expression $\widetilde{\omega}_\mathrm{s\mbox{-}o}=\sqrt{c_{\Phi}^2+s_{\Phi}^2 p^2}$.

\section{Concluding remarks}
As a magnetized object is driven by an externally applied magnetic field, the equations governing its evolution are invariant only under symmetries which preserve this field. The in-plane rotating magnetic field ${\bm H}=H(\hat{\bm x}\cos \omega t+\hat{\bm y}\sin \omega t)$ is invariant under parity $\widehat{P}$ and $\widehat{C}\widehat{R}_z$ that involves charge conjugation. It was demonstrated in \cite{pre18} that highly symmetrical (achiral, i.e, $\widehat{P}$-even) planar V-shaped objects exhibit no net propulsion while individual less symmetrical ($\widehat{C}\widehat{P}$-even) propellers can propel quite efficiently. In the latter case the propulsion direction, i.e., $+z$ or $-z$, is controlled by the object initial orientation which serves as to `spontaneously break the symmetry', thus a large collection of such propellers having
random initial orientations would at most exhibit symmetric spreading with zero ensemble average velocity. This finding is relevant to practical applications, as it indicates that a collection of 2D ferromagnetic micro/nanomotors that are prone to in-plan magnetization and can be fabricated using standard lithography methods, could not be steered in a controlled fashion. Particular orientation of the magnetic moment, $\bm m$, rendering the V-shape $\widehat{C}\widehat{P}$-chiral does yield unidirectional propulsion typically associated with helical structures, however, it requires off-plane magnetization which is not easy to achieve.

It was pointed out in \cite{pre18} that $\widehat{C}\widehat{R}_z$-symmetry is special to the case of \emph{plane} rotating field and it would be interesting to explore modification of the actuating field that breaks this symmetry. In the present paper we examined how the results of \cite{pre18} change upon adding a constant magnetic field along the field rotation $z$-axis that breaks the $\widehat{R}_z\widehat{C}$-symmetry, but preserves $\widehat{P}$-symmetry. Our analysis confirms that such modification of the actuating field removes the degeneracy of the plane rotating field and results in enantiomeric selection of the propulsion direction of magnetized in-plane symmetric V-shaped objects (see Fig.~\ref{fig:U2}). Surprisingly, magnetization along the V-shape symmetry axis (rendering the 2D object ($\widehat{P}$)-chiral) results in unidirectional in-sync propulsion with  \emph{constant} (frequency-independent) speed in a limited range of frequencies (see Fig.~\ref{fig:U}). Recall that for an in-plane rotating magnetic field, a magnetization along any principal axis yielded no propulsion. The $\widehat{P}$-even object magnetized along a principal axis still exhibits no propulsion even when actuated by this less symmetrical conically rotating field.

Note that for highly symmetrical ($\widehat{C}\widehat{P}$-achiral) V-shaped propellers the orientation of the constant field controls the direction of propulsion, similar to what was observed for flexible magnetic nanowires in \cite{Pak}. For instance, the direction of propulsion in Fig.~\ref{fig:U} will not change upon reversal of the direction of the field rotation. The results shown in Fig.~\ref{fig:U2} demonstrate that the same holds more generally provided that $\delta$ is not too small. For rigid magnetic helices, on the other hand, the propulsion direction is controlled by the rotation direction of the magnetic field and its intrinsic handedness, while reversal of propulsion is anticipated upon reversal of the field rotation. The flexible nanowires in \cite{Pak} preserved the propulsion direction upon reversal of the field rotation because they had no intrinsic handedness and the acquired chirality was controlled by the direction of the field rotation.

The developed theory of magnetic actuation using a conically rotating field should be most relevant towards enhanced control of propulsion of swarms of 2D micro/nanopropellers without the need of individual-level feedback control \cite{ASME14}.

\section*{Acknowledgement}
This work was supported in part by the Israel Science Foundation (ISF) via the grant No. 1744/17 (A.M.L.) 
The authors wish to thank Johannes Sachs and Peer Fischer for fruitful discussions.

\begin{appendices}

\section{Propulsion of a symmetric V-shaped propeller \label{A}}
\setcounter{equation}{0}
\renewcommand{\theequation}{A\arabic{equation}}

In the laboratory frame, the translational velocity of a propeller is $\bm U^l={\bm R}^T \cdot \bm U^b$ where ${\bm R}^T$ is the transposed rotation matrix.
From Eqs.~(\ref{eq:U}) we have $ \bm U^b={\bcalG}\cdot \bm{L}^b={\bcalG}\cdot{\bcalF}^{-1}\cdot \bm{\mathit{\Omega}}^b$ where the components of the angular velocity $\bm{\mathit{\Omega}}^b$ in the body-frame are determined by Eq.~(\ref{eq:Omega}). For the chubby V-shape propeller (as in Fig.~\ref{fig:schematic}b), the only non-trivial entries of the coupling matrix ${\bcalG}$ are ${\mathcal G}_{13}\equiv{\mathcal G}_{31}$. Therefore the linear velocity in the body-frame reads
\begin{equation}
\bm U^b= \frac{{\mathcal G}_{13}}{{\mathcal F}_{3}} {\mathit{\Omega}}^b_3 {\bm e}_1+\frac{{\mathcal G}_{31}}{{\mathcal F}_{1}}
{\mathit{\Omega}}^b_1 \bm{e}_3\,.
\label{eq:Ub}
\end{equation}
Thus, the components of the translational velocity in the laboratory frame read:
\begin{eqnarray}
U_x&=&\textstyle\frac {{\mathcal G}_{13}}{{\mathcal F}_{3}}(c_{\varphi}c_{\psi}-s_{\varphi}s_{\psi}c_{\theta}){\mathit{\Omega}}^b_3+
\frac {{\mathcal G}_{31}}{{\mathcal F}_{1}}s_{\varphi}s_{\theta}{\mathit{\Omega}}^b_1\,,\nonumber\\
U_y&=&\textstyle\frac {{\mathcal G}_{13}}{{\mathcal F}_{3}}(s_{\varphi}c_{\psi}+c_{\varphi}s_{\psi}c_{\theta}){\mathit{\Omega}}^b_3-
\frac {{\mathcal G}_{31}}{{\mathcal F}_{1}}c_{\varphi}s_{\theta}{\mathit{\Omega}}^b_1\,,\label{eq:Ub2}\\
U_z&=&\textstyle\frac {{\mathcal G}_{13}}{{\mathcal F}_{3}}s_{\psi}s_{\theta}{\mathit{\Omega}}^b_3+
\frac {{\mathcal G}_{31}}{{\mathcal F}_{1}}c_{\theta}{\mathit{\Omega}}^b_1\,.\nonumber
\end{eqnarray}
It is seen that in the in-sync regime where $\varphi=\omega t+\mathrm{Const}$, $\psi=\mathrm{Const}$, $\theta=\mathrm{Const}$, the components $U_x$ and $U_y$ oscillate with the
field frequency $\omega$ and have zero mean upon averaging over a period $T=2\pi/\omega$. At the same time, the component $U_z$ does not depend on time.

Let us consider the propulsion velocity $U_z$  of the symmetric V-shape. Substituting the components of
the angular velocity (\ref{eq:Omega}) into the third equation of Eqs.~(\ref{eq:Ub2}) one readily finds
\begin{equation}
\frac{U_z}{\ell}=\widetilde{\mathrm {Ch}}s_{\psi}s_{2\theta}\dot{\varphi}+
\frac {{\mathcal G}_{13}}{{\mathcal F}_{3}}s_{\psi}s_{\theta}\dot{\psi}+
\frac {{\mathcal G}_{31}}{{\mathcal F}_{1}}c_{\psi}c_{\theta}\dot{\theta}\,,
\label{eq:U4}
\end{equation}
where $\widetilde{\mathrm {Ch}}=\mathcal{G}_{13}({{\cal F}_{1}}^{-1}+{{\cal F}_{3}}^{-1})/2\ell$.

\end{appendices}


\begin{thebibliography}{99}
\bibitem{note0} Nondimensionalization with $\nu_0$ complies with the definition of the propulsion efficiency $\delta^*= |U_{z,\,\mathrm{s\mbox{-}o}}|/\nu_0\ell$ introduced in Ref.~\cite{scirob} that ranks propellers according to their maximal speed (at the step-out) in body lengths per unit time.
\bibitem{note3} Sometimes a different limit cycle solution can be obtained, however its basin of attraction is rather narrow, so for most initial orientations the solution converges to the same limit cycle.
\bibitem{note2} The condition $qr>s$ follows from the fact that the roots of the cubic equation satisfy $(\widetilde{\lambda}_1+\widetilde{\lambda}_2) (\widetilde{\lambda}_1+\widetilde{\lambda}_3) (\widetilde{\lambda_2}+\widetilde{\lambda}_3)=s-qr$.
\bibitem{note1} For a slim V-structure (as in Fig.~\ref{fig:schematic}a) magnetization along $\bm e_1$ results in propulsion similar to that in Sec.~\ref{sec:e2} subject to
some changes due to axes interchange, $\bm e_2 \leftrightarrow \bm e_1$. The in-sync propuslion velocity in this case is given by (\ref{eq:U1}) with $\psi \rightarrow \pi/2-\psi$, $\theta \rightarrow \pi-\theta$ and $\widetilde{\mathrm {Ch}}=\mathrm {Ch}_{23}$, resulting in
$U_z/\omega \ell=- \widetilde{\mathrm {Ch}}\, 2\delta/(1+\varepsilon - p^{-1})$. Since $\widetilde{\mathrm {Ch}}$ changes sign, the propulsion in the $+z$-direction will require magnetization along $-\bm e_1$.
\end{thebibliography}
\end{document}